\documentclass[12pt]{article}
\usepackage{bbold}
\usepackage{amsmath}
\newcommand{\ket}[1]{| #1 \rangle}
\begin{document}
\date{}
\title{Quantum Mechanics and Perspectivalism}
\author{Dennis Dieks \\History and Philosophy of Science\\
Utrecht University\\ d.dieks@uu.nl}
\maketitle
\begin{abstract}
Experimental evidence of the last decades has made the status of ``collapses of the wave function'' even more shaky than it already was on conceptual grounds: interference effects turn out to be detectable even when collapses are typically expected to occur. Non-collapse interpretations should consequently be taken seriously. In this paper we argue that such interpretations suggest a \emph{perspectivalism} according to which quantum objects are not characterized by monadic properties, but by \emph{relations} to other systems. Accordingly, physical systems may possess different properties with respect to different ``reference systems''. We discuss some of the relevant arguments, and argue that perspectivalism both evades recent arguments that single-world interpretations are inconsistent and eliminates the need for a privileged rest frame in the relativistic case. 
\end{abstract}

\section{Introduction: early hints of non-collapse and perspectivalism}\label{intro}

In introductions to quantum mechanics it is standard to introduce ``collapses of the wave function'' in order to avoid the occurrence of superpositions of states associated with different macroscopic properties. The paradigm case is the quantum mechanical treatment of measurement: if the interaction between a quantum system and a measuring device is described by means of unitary Schr\"{o}dinger evolution, the composite system of object plus device will generally end up in an entangled state that is not an eigenstate of the measured observable, but rather a superposition of such states. But successful measurements end with the realization of exactly one of the possible outcomes, so it appears plausible that at some stage during the measurement interaction unitary evolution is suspended and a \emph{collapse} to one of the terms in the superposition takes place.  

However, experimental research of the last decades has undermined this motivation for the introduction of collapses. ``Schr\"{o}dinger cat states'', i.e.\ superpositions of distinguishable quantum states of mesoscopic or even practically macroscopic physical systems are now routinely prepared in the laboratory, and interference between the different terms in the superpositions have abundantly been verified (see e.g.\ \cite{johnson} for a sample of recent developments).  This lends inductive support to the hypothesis that superpositions never really collapse, but are merely difficult to detect in everyday situations. In such situations  huge numbers of environmental degrees of freedom come into play, so that the mechanism  of decoherence may be invoked as an explanation for the practical unobservability of interference between macroscopically different states under standard conditions.  This line of thought leads in the direction of non-collapse interpretations of quantum mechanics. 

The evidence against collapses has not yet affected the textbook tradition, which has not questioned the status of collapses as a mechanism of evolution alongside unitary Schr\"{o}dinger dynamics. However, the relevant views of the pioneers of quantum mechanics were not at all clear-cut. The \emph{locus classicus} for the introduction and discussion of collapses is chapter 6 of Von Neumann's 1932 \emph{Mathematical Foundations of Quantum Mechanics} \cite{neumann}. In this chapter Von Neumann underlines the fundamental difference between collapses---occurring in measurements---and unitary evolution, but connects this difference to the distinction between on the one hand the experience of an observer and on the other hand external descriptions (in which the observer is treated in the same way as the other physical systems involved in the measurement interaction).  In the external description Von Neumann assumes unitary evolution, with superpositions (also involving the observer) as an inevitable consequence. Nevertheless, Von Neumann states, the content of the observer's ``subjective experience'' corresponds to only one single term in the superposition. 

So the distinction between collapses and unitary evolution for Von Neumann is not a distinction between two competing and potentially conflicting physical interaction mechanisms on the same level of description, but rather concerns what can be said \emph{in relation to two different points of view}---an idea taken up and developed by London and Bauer (see below).    

Also Niels Bohr took the position that the standard rules of quantum mechanics apply even to measuring devices and other macroscopic objects, so that strictly speaking these cannot be characterized by sets of precise values of classical quantities (even though folklore has it that Bohr assumed that quantum mechanics does not apply to the macroscopic world, see \cite{dieksbohr} for an extensive discussion).   Thus, in 1935 Bohr wrote \cite{bohrepr} that ``a purely classical account of the measuring apparatus still implies the necessity of latitudes corresponding to the uncertainty relations. If spatial dimensions and time intervals are sufficiently large, this involves no limitation.'' And in 1948 \cite{bohrdialectica} he commented in the same vein: ``we may to a very high degree of approximation disregard the quantum character of the measuring instruments if they are sufficiently heavy.'' Although for Bohr there is thus no difference of principle between macro and micro objects, he does assign a special role to the observer and to the ``conditions of measurement'' (the specific experimental set-up, chosen by the observer). However, he does not acknowledge a \emph{sui generis} measurement dynamics but rather refers to the specific epistemological vantage point of the observer, who can only communicate what he finds by using definite values of classical quantities (the paradigm case being the assignment of a definite value of either position or momentum, depending on the chosen kind of measuring device). 

According to Bohr, the object that is being measured and the measuring device form in each individual case one insoluble whole,  so that ``an independent reality in the ordinary physical sense can not be ascribed to the phenomena'' \cite{bohr1928}. The properties of a quantum system according to Bohr only become well-defined in the context of the system's coupling to a measuring device---which points in the direction of a relational nature of physical properties. 

A more formal analysis of quantum measurements, close to Von Neumann's account, was given by London and Bauer in their 1939 booklet on the \emph{Theory of Observation in Quantum Mechanics} \cite{london}.  London and Bauer consider three interacting systems: $\mathbf{x}$, the object system, $\mathbf{y}$, a measuring device, and $\mathbf{z}$, the observer. As a result of the unitary evolution of the combined object--device system, an entangled state will result: $ \Sigma_k c_k |x \rangle_k |y \rangle_k    $.  When the observer reads off the result of the measurement, a similar unitary evolution of the $\mathbf{x}, \mathbf{y}, \mathbf{z}$ system takes place, so that the final state becomes: $ | \Psi \rangle = \Sigma_k c_k |x \rangle_k |y \rangle_k   |z \rangle_k $. London and Bauer comment \cite[pp.\ 41--42]{london}:
\begin{quote}
``Objectively''---that is, for us who consider as ``object'' the combined system $\mathbf{x}, \mathbf{y}, \mathbf{z}$---the situation seems little changed compared to what we just met when we were only considering apparatus and object. ... The observer has a completely different viewpoint: for him it is only the object $\mathbf{x}$ and the apparatus $ \mathbf{y}$ which belong to the external world, to that what he calls ``objective''. By contrast, he has \emph{with himself} relations of a very special character: he has at his disposal a characteristic and quite familiar faculty which we can call the ``faculty of introspection.'' For he can immediately give an account of his own state. By virtue of this ``immanent knowledge'' he attributes to himself the right to create his own objectivity, namely, to cut the chain of statistical correlations expressed by $ \Sigma_k c_k |x \rangle_k |y \rangle_k   |z \rangle_k $ by stating ``I am in the state $|z \rangle_k $'', or more simply ``I see $y_k$'' or even directly ``$X = y_k$''\footnote{Here $X$ stand for the observable whose value is measured by the apparatus.}.
\end{quote}

It is clear from this quote and the further context that London and Bauer believed that there is a role for human consciousness in bringing about a definite measurement outcome---even though they also assumed, like von Neumann, that ``from the outside'' the observer, including his consciousness, can be described in a physicalist way, by unitary quantum evolution (cf.\ \cite[sect.\ 11.3]{jammer}).  The appeal to consciousness can hardly be considered satisfactory, though: it appears to invoke a \emph{deus ex machina}, devised for the express purpose of reconciling unitary evolution with definite measurement results. More generally, the hypothesis that the definiteness of the physical world only arises as the result of the intervention of (human?) consciousness does not sit well with the method of physics. 

Although certain elements of London's and Bauer's solution are therefore hard to accept, the suggestion that it should somehow be possible to reconcile universal unitary evolution, and the resulting omnipresence of entangled states, with the occurrence of definite values of physical quantities appears plausible. Indeed, the theoretical framework of quantum mechanics itself\footnote{As opposed to modifications of the quantum formalism, as in the GRW-theory.} does not in a natural way leave room for another dynamical process beside unitary evolution; e.g., there is no time scale or scale of complexity at which this alternative evolution could set in. As already mentioned, empirical results support this verdict. Accordingly, in the next sections we shall investigate whether the early intuitions about the universality of unitary evolution, excluding collapse as a dynamical process, can be salvaged in a purely physicalist way. We shall argue that ``perspectival'' non-collapse interpretations capture the intuitions behind the London and Bauer and Von Neumann analyses, without an appeal to consciousness or human observers.  

\section{Relational Aspects of Non-Collapse Interpretations}\label{noncollapse}

The common feature of non-collapse interpretations is that they single out unitary evolution (Schr\"{o}dinger evolution or one of its relativistic generalizations) as the only way that quantum states develop in time. Consequently, entangled quantum states generally develop as a result of interactions, even with macroscopic objects like measuring devices. Of course, the task is to reconcile this with the definite states of affairs encountered in experience.

There are several proposals for such a reconciliation. The best-known one is probably the many-worlds interpretation, according to which each individual term in a superposition that results from a measurement-like interaction represents an actual state of affairs, characterized by definite values of some set of observables. In this approach there are many actual states of affairs, worlds or ``branches'', living together in a ``super-universe''. The experience of any individual observer is restricted to one single branch within this super-universe. In other words, the experienced world is the part of the super-universe that is accessible from the observer's perspective (a relational aspect of the scheme, which is the reason that Everett first introduced it as the \emph{``relative state'' formulation of quantum mechanics} \cite{everett}). 

A second category of interpretations, modal interpretations, holds that there is only \emph{one} actual reality, so that all except one of the ``branches''of the total entangled state do not correspond to actual worlds but rather to \emph{unrealized possibilities---``modalities''}. Some of these modal interpretations make the assumption that there is one \emph{a priori} preferred observable (or set of commuting observables) that is always definite-valued in each physical system, others assume that the set of definite-valued quantities depends on the form of the quantum state and can therefore change over time (see \cite{bub,dieksvermaas,lombardidieks} for overviews). In the first category belong the Bohm interpretation, in which \emph{position} is always definite, and the ``Hamiltonian modal interpretation'' \cite{lombardi} according to which \emph{energy} plays a privileged role. An example of the second category is the proposal according to which the bi-orthogonal (Schmidt) decomposition of the total state determines the definite quantities of partial systems (namely, the quantities represented by the projection operators projecting on the basis vectors that diagonalize the partial system's density matrix \cite{vermaas}); another proposal is to make decoherence responsible for the selection of definite quantities, in the same way as is now standard in many-worlds accounts. 

Other non-collapse approaches are the consistent-histories interpretation \cite{griffiths} and Rovelli's relational interpretation \cite{rovelli,laudisa}. Interestingly, the latter interpretation posits from the outset that the dynamical properties of any physical system are purely \emph{relational} and only become definite \emph{with respect to some other system} when an interaction between the two systems (in the formalism described by unitary quantum evolution) correlates the systems (so that there is an ``exchange of information'' between them).  
       
However, relational features also have a natural place in most of the other just-mentioned non-collapse interpretations (although not in all of them), as can be illustrated by further considering the situation discussed by London and Bauer (section \ref{intro})---which essentially is the well-known ``Wigner's friend'' thought experiment.  

Suppose that an experimentalist (our friend, who is a perfect observer) performs a quantum measurement within a hermetically sealed room. 
Let us say that the spin of a spin-$1/2$ particle is measured in a previously fixed direction, and that the experimentalist notes the outcome (either $+1/2$ or $-1/2$).  After some time we, who are outside the room, will be sure that the experiment is over and that our friend will have observed a definite result. Yet, we possess no certainty about the outcome. In a classical context we would therefore represent the state of the room and its contents by an ignorance mixture over states: there are two possibilities (``up'' and ``down''), both with probability $1/2$. 

But in unitary quantum mechanics the situation is different in an important respect. According to the Von Neumann measurement scheme, the final situation of the room after the experiment, including a record of the friend's observation, will be given by a linear superposition of terms, each containing a definite spin state of the particle coupled to a state of our experimentalist in which he is aware of his found spin value. For us outside, this superposition is the correct theoretical description of the room and its contents; and this (coherent) superposition is different from an (incoherent) ignorance mixture over different possible states. As mentioned in the previous section, experience supports the ascription of this superposed state: experiments with Schr\"{o}dinger cat states demonstrate that we need the superposition to do justice to the experimental facts. For example, if we are going to measure the projection operator $|\Psi\rangle \langle \Psi |$ (where $|\Psi\rangle $ stands for the superposed state of the room and its contents), the formalism tells us that we shall find the result ``$1$'' with certainty; this is different from what a mixed state would predict. Experiment confirms predictions of this kind.

However, we also possess robust experience about what happens when we watch an experiment inside a closed laboratory room: there will be a definite outcome. Therefore, it seems inevitable to accept that during the experiment our friend becomes aware of exactly one spin value. As stated by London and Bauer, our friend will be justified in saying either ``the spin is up'' or ``the spin is down'' after the experiment.  The dilemma is that we, on the outside, can only derive an ``improper mixture'' as a state for the particle spin, and that well-known arguments forbid us to think that this mixture represents our ignorance about the actually realized spin-eigenstate (indeed, if the spin state actually was one of the up or down eigenstates, it would follow that the total system of room and its contents had to be an ignorance mixture as well, which conflicts with the premise---supported both theoretically and empirically---that the total state is a superposition).

Our proposed perspectival way out of this dilemma is to ascribe more than one state to the same physical system. In the case under discussion, with respect to us, representing the outside point of view, the contents of the laboratory room are correctly described by an entangled pure state so that we should ascribe improper mixtures (obtained by ``partial tracing'') to the inside observer, the measuring device and the spin particle. But with respect to the inside observer (or with respect to the measuring device in the room) the particle spin is definite-valued. So the inside observer should assign a state to his environment that appropriately reflects this definiteness. 

This line of thought leads to the idea of assigning \emph{relational} or \emph{perspectival} states, i.e.\ states of a physical system $A$ \emph{from the perspective} of a physical system $B$. This step creates room for the possibility that the state and physical properties of a system $A$ are different in relation to different ``reference systems'' $B$. As suggested by the examples, this move may make it possible to reconcile the unitary evolution during a quantum measurement with the occurrence of definite outcomes. The properties associated with the superposition and the definite outcomes, respectively, would relate to two different perspectives---the idea already suggested by Von Neumann and London and Bauer. Of course, we should avoid the earlier problems associated with consciousness and ill-defined transitions. The different perspectives, and different relational states, should therefore be defined in purely physical terms.

The idea as just formulated was tentative: we spoke in a loose way of ``states'', thinking of wave functions (or vectors in Hilbert space) without specifying what the attribution of quantum states to physical systems means on the level of physical quantities, i.e.\ in terms of physical \emph{properties} of the systems concerned. In fact, this physical meaning is interpretation-dependent. 

In the many-worlds interpretation the perspectival character of quantum states, for the Wigner's Friend type of scenario that we just discussed, translates into the following physical account. When the measurement interactions within the hermetically sealed room have completely ended, the contents in the room have split into two copies: one in which the outcome $+1/2$ has been realized and observed, and one with the outcome $-1/2$. However, we as external observers can still verify the superposed state by measuring an observable like $| \Psi \rangle \langle \Psi |$, so that for us the two ``worlds'' inside the room still form one whole. Apparently, the splitting (branching) of worlds that happens in measurements cannot be a global process, extending over the whole universe at once, but must be a local splitting that propagates with the further physical interactions that take place (cf.\ \cite{bacciagaluppi}). Therefore,  although we know (if we reason in terms of the many-worlds interpretation) that there are two copies of our friend inside, each having observed one particular outcome, we still consider the room plus its contents as represented by the coherent superposition that corresponds to the definite value ``$1$'' of the physical quantity represented by the observable $| \Psi \rangle \langle \Psi |$. So here we encounter a perspectivalism on the level of physical properties: there exists a definite spin value for the internal observer, but not for his external colleague.

The same type of story can be told in those modal interpretations in which the definite-valued physical properties of systems are defined by their quantum states (one detailed proposal for how to define physical properties from the quantum state can be found in  \cite{benedieks,dieksrel}). The main difference with the many-worlds account is that now the interactions within the room do not lead to \emph{two} worlds but to only \emph{one}, with either spin up or spin down. Also in this case, there is a definite spin value for the internal observer whereas from the outside it is rather the observable $| \Psi \rangle \langle \Psi |$ (and observables commuting with it) that is definite-valued, which conflicts with the attribution of a value to the spin---even though outside observers may know that for their inside counterpart there is such a value.

Rovelli's relational interpretation, which takes part of its inspiration from Heisenberg's heuristics in the early days of quantum mechanics, says that a quantity of system $B$ only becomes definite for $A$ when an interaction (a measurement) occurs between $A$ and $B$ \cite{rovelli}. In our Wigner's friend scenario, this again leads to the verdict that the internal interactions in the laboratory room lead to a situation in which the spin is definite with respect to an internal observer, but not for an external one. Only when (and if) external observers enter the room and interact with the spin system, does the spin become definite for them as well.

In all these cases we obtain accounts that are similar to the London and Bauer analysis, but with the important distinction that non-physical features do not enter the story. It should be noted that the relational properties introduced here are intended to possess an ontological status: it is not the case that for an outside observer the internal spin values are definite but unknown. The proposal is that the spin really \emph{is} indeterminate with respect to the world outside the laboratory room.

This perspectivalism with respect to properties does not seem an inevitable feature of all non-collapse interpretations, however. In particular, those interpretations of quantum mechanics in which it is assumed that there exists an \emph{a priori} given set of  preferred observables that is always definite---in all physical systems, at all times and in all circumstances---are by construction at odds  with the introduction of a definiteness that is merely relative. The Bohm interpretation is a case in point. According to this interpretation all physical systems are composed of particles that always possess a definite \emph{position}, as a monadic attribute independent of any perspective. So in our sealed room experiment the instantaneous situation inside is characterized by the positions of all particles in the room, and this description is also valid with respect to the outside world---even though an outside observer will usually lack information about the exact values of the positions. So for an external observer there exists one definite outcome of the experiment inside, corresponding to one definite particle configuration. The outcome of any measurement on the room as a whole that the outside observer might perform again corresponds to a definite configuration of particles with well-defined positions. The fact that this value is not what we would classically expect (for example, when we measure $| \Psi \rangle \langle \Psi |$) is explained by the Bohm theory via the non-classical measurement interaction between the external observer's measuring device and the room. The quantum states that in perspectival schemes encode information about which \emph{physical properties} are definite, in the Bohm types of interpretations only play a role in the \emph{dynamics} of a \emph{fixed} set of quantities, so that the possibility of relational properties or perspectivalism does not suggest itself.

However, it has recently been argued that all interpretations of this unitary kind, characterized by definite and unique outcomes at the end of each successful experiment even though the total quantum state always evolves unitarily, cannot be consistent \cite{frauchiger,araujo,bub2}. This argument is relevant for our theme, and we shall discuss it in some detail.

\section{Unitarity and Consistency}\label{context}

In a recent paper \cite{frauchiger} Frauchiger and Renner consider a sophisticated version of the Wigner's friend experiment in which there are two friends, each in their own room, with a private information channel between them. Outside the two rooms are Wigner and an Assistant. The experiment consists of a series of four measurements, performed by the individual friends, and then by the Assistant and Wigner, respectively. In the room of Friend 1 a quantum coin has been prepared in a superposition of ``heads'' and ``tails'': $\frac{1}{\sqrt{3}}\ket{h} + \sqrt{\frac{2}{3}}\ket{t}$. The experiment starts when Friend 1 measures her coin, and finds either heads (probability $1/3$) or tails (probability $2/3$). Friend 1 then prepares a qubit in the state $\ket{0}$ if her outcome is $h$, and in the state  $\frac{1}{\sqrt{2}}(\ket{0} + \ket{1})$ if her outcome is $t$, and sends the qubit via the private channel between the rooms to Friend 2. When Friend 2 receives the qubit, he subjects it to a measurement of an observable that has the eigenstates $\ket{0}$ and $\ket{1}$. As in the thought experiment of section \ref{noncollapse}, the external observers subsequently measure ``global'' observables on the respective rooms; this is first done by the Assistant (on the room of Friend 1), and then by Wigner (on the room of Friend 2).

Frauchiger and Renner claim, via a rather complicated line of reasoning\footnote{M. Ara\'{u}jo has given a concise version of the argument \cite{araujo}, which makes the ideas clearer.}, that any interpretation of quantum mechanics that assigns unique outcomes to these meaurements ``in one single world'', while using only unitary evolution for the dynamics of the quantum state (also during the measurements), will lead to an inconsistent assignment of values to the measurement outcomes. If this conclusion is correct, there are significant implications for the question of which unitary interpretations of quantum mechanics are possible. The theories that are excluded according to Frauchiger and Renner are theories that ``rule out the occurrence of more than one single outcome if an experimenter measures a system once'' \cite[p.\ 2]{frauchiger}. If this were right, accepting many worlds would seem inevitable.  In fact, Frauchiger and Renner themselves conclude that ``the result proved here forces us to reject a single-world description of physical reality'' \cite[p.\ 3]{frauchiger}. 

However, we should not be too quick when we interprete this statement. As Frauchiger and Renner make clear, they use their ``single-world assumption'' to ensure that all measurement outcomes are \emph{context-independent}. In particular, what they use in their proof is a compatibility condition between different ``stories'' of a measurement: if one experimenter's story is that an experiment has outcome $t$, this should entail that in every other experimenter's story of the same event this same outcome $t$ also figures \cite[p.\ 7]{frauchiger}. This is first of all a denial of the possibility of perspectivalism. As we shall further discuss in a moment, perspectival interpretations will be able to escape the conclusion of the F-R argument. Therefore, what we are going to claim is that the F-R argument can be taken to lend support to perspectivalism, as one of the remaining consistent possibilities.

The details, and domain of validity, of the F-R proof are not completely transparent and uncontroversial. Indeed, there is at least one non-perspectival single-world interpretation, namely the Bohm interpretation, whose consistency is usually taken for granted. This consistency is confirmed by a result of Sudbery \cite{sudbery}, who has concretely constructed a series of outcomes for the F-R thought experiment as predicted by a modal interpretation of the Bohm type. According to Sudbery there is an unjustified step in Frauchiger's and Renner's reasoning, because they do not fully take into account that in unitary interpretations only the total (non-collapsed) state can be used for predicting the probabilities of results obtained by the Assistant and Wigner\footnote{The bone of contention is  statement 4 in Ara\'{u}o's reconstruction of the F-R inconsistency \cite[p.\ 4]{araujo}, in which Friend 1 argues that his coin measurement result is only compatible with one single later result to be obtained by Wigner in the final measurement. But in unitary interpretations previous measurement results do not always play a role in the computation of probabilities for future events. Indeed, a calculation on the basis of the total uncollapsed quantum state as given on p.\ 4 of \cite{araujo} indicates that Wigner may find either one of two possible outcomes, with equal probabilities, even given the previous result of Friend 1---this contradicts the assumption made by Frauchiger and Renner.}.

The situation becomes much less opaque when we make use of an elegant version of the F-R thought experiment recently proposed by Bub \cite{bub2}. Bub replaces Friend 1 by Alice and Friend 2 by Bob; Alice and Bob find themselves at a great distance from each other.  Alice has a quantum coin which she subjects to a measurement of the observable $A$ with eigenstates $\ket{h}_{A}, \ket{t}_{A}$; as before, the coin has been prepared in the initial state   $\frac{1}{\sqrt{3}}\ket{h}_{A} + \sqrt{\frac{2}{3}}\ket{t}_{A}$. Alice then prepares a qubit in the state $\ket{0}_{B}$ if her outcome is $h$, and in the state  $\frac{1}{\sqrt{2}}(\ket{0}_{B} + \ket{1}_{B})$ if her outcome is $t$. She subsequently sends this qubit to Bob---this is the only ``interaction'' between Alice and Bob. After Bob has received the qubit, he subjects it to a measurement of the observable $B$ with eigenstates $\ket{0}_{B}, \ket{1}_{B}$. 

In accordance with the philosophy of non-collapse interpretations, we assume that Alice and Bob obtain definite outcomes for their measurements, but that the total system of Alice, Bob, their devices and environments, and the coin and the qubit can nevertheless be described by the uncollapsed quantum state, namely: 
\begin{equation}
\ket{\Psi}  =  \frac{1}{\sqrt{3}}(\ket{h}_{A}\ket{0}_{B} + \ket{t}_{A}\ket{0}_{B} +\ket{t}_{A}\ket{1}_{B}). 
\end{equation}
For ease of notation the quantum states of Alice and Bob themselves, plus the measuring devices used by them, and even the states of the environments that have become correlated to them, have here all been included in the states $\ket{h}_{A}$, $\ket{t}_{A}$, $\ket{0}_{B})$ and $\ket{1}_{B}$ (so that these states no longer simply refer to the coin and the qubit, respectively, but to extremely complicated many-particles systems!).

Now we consider two external observers, also at a very great distance from each other, who take over the roles of Wigner and his assistant, and are going to perform measurements on Alice and Bob (and their entire experimental set-ups), respectively. The external observer who focuses on Alice measures an observable $X$ with eigenstates $\ket{\mbox{fail}}_{A} =  \frac{1}{\sqrt{2}}(\ket{h}_{A} + \ket{t}_{A})$ and $\ket{\mbox{ok}}_{A} = \frac{1}{\sqrt{2}}(\ket{h}_{A} - \ket{t}_{A})$, and the observer dealing with Bob and Bob's entire experiment measures the observable $Y$ with eigenstates $\ket{\mbox{fail}}_{B} =  \frac{1}{\sqrt{2}}(\ket{0}_{B} + \ket{1}_{B})$ and $\ket{\mbox{ok}}_{B} =  \frac{1}{\sqrt{2}}(\ket{0}_{B} - \ket{1}_{B})$. 

A F-R contradiction now arises in the following manner \cite[p.\ 3]{bub2}. The state $\ket{\Psi}$ can alternatively be expressed as:
\begin{eqnarray}
\ket{\Psi} & = & \frac{1}{\sqrt{12}}\ket{\mbox{ok}}_{A}\ket{\mbox{ok}}_{B} - \frac{1}{\sqrt{12}}\ket{\mbox{ok}}_{A}\ket{\mbox{fail}}_{B} \nonumber\\
&& + \frac{1}{\sqrt{12}}\ket{\mbox{fail}}_{A}\ket{\mbox{ok}}_{B} + \sqrt{\frac{3}{4}}\ket{\mbox{fail}}_{A}\ket{\mbox{fail}}_{B}\\
& = & \sqrt{\frac{2}{3}}\ket{\mbox{fail}}_{A}\ket{0}_{B} + \frac{1}{\sqrt{3}}\ket{t}_{A}\ket{1}_{B}\\
& = & \frac{1}{\sqrt{3}}\ket{h}_{A}\ket{0}_{B} + \sqrt{\frac{2}{3}}\ket{t}_{A}\ket{\mbox{fail}}_{B}
\end{eqnarray}

From Eq.\ 2, we see that the outcome $\{\mbox{ok, ok}\}$ in a joint measurement of $X$ and $Y$ has a non-zero probability: this outcome will be realized in roughly $1/12$-th of all cases if the experiments are repeated many times. From Eq.\ 3 we calculate that the pair $\{\mbox{ok}, 0\}$ has zero probability as a measurement outcome, so $\{\mbox{ok}, 1\}$ is the only possible pair of values for the observables $X, B$ in the cases in which $X$ has the value $\mbox{ok}$. However, from Eq.\ 4 we conclude that the pair $\{t, \mbox{ok}\}$ has zero probability, so $h$ is the only possible value for the observable $A$ if $Y$ has the value $\mbox{ok}$ and $A$ and $Y$ are measured together. So this would apparently lead to the pair of values $\{h, 1\}$ as the only possibility for the observables $A$ and $B$, if $X$ and $Y$ are jointly measured with the result $\{\mbox{ok, ok}\}$. But this pair of values has zero probability in the state $\ket{\Psi}$, so it is not a possible pair of measurement outcomes for Alice and Bob in that state. So although the outcome $\{\mbox{ok, ok}\}$ for $X$ and $Y$ is certainly possible, the (seemingly) necessarily associated outcome $\{h, 1\}$ for $A$ and $B$ is not---this is an inconsistency. 

In this inconsistency argument there is a silent use of the F-R non-perspectivalism condition. For example, it is assumed that if Bob's measurement outcome is $1$ from the perspective of the $X$ measurement, this outcome has also to be $1$ as judged from the perspective of the $Y$ observer. But it is not evident that this assumption sits well with the quantum formulas: for example, the relative state of Bob with respect to the $Y$ outcome ``ok''  is not $\ket{1}_B$, but $\ket{\mbox{ok}}_B$ (cf.\ Eq.\ 5 below), and it is not immediately clear that before the $Y$ measurement it had to be $1$. 

To see in more detail what is wrong with the inconsistency argument from a viewpoint that closely follows the quantum formalism, it is helpful to note  that the states (1), (2), (3) and (4) are all states of Alice and Bob, including their devices and environments, but without the external observers. In a consistent non-collapse interpretation we must also include the external observer states in the total state if we want to discuss the measurements of the observables $X$ and $Y$.  If we denote by $\ket{\mbox{o}}$ and $\ket{\mbox{f}}$ the external states corresponding to the measurement results ``ok'' and ``fail'', respectively, we find for the final state, in obvious notation:
\begin{eqnarray}
 && \frac{1}{\sqrt{12}}\ket{\mbox{ok}}_{A}\ket{\mbox{ok}}_{B}\ket{\mbox{o}}_{X}\ket{\mbox{o}}_{Y} - \frac{1}{\sqrt{12}}\ket{\mbox{ok}}_{A}\ket{\mbox{fail}}_{B}\ket{\mbox{o}}_{X}\ket{\mbox{f}}_{Y} \nonumber \\ && + \frac{1}{\sqrt{12}}\ket{\mbox{fail}}_{A}\ket{\mbox{ok}}_{B}\ket{\mbox{f}}_X\ket{\mbox{o}}_Y + \sqrt{\frac{3}{4}}\ket{\mbox{fail}}_{A}\ket{\mbox{fail}}_{B}\ket{\mbox{f}}_X\ket{\mbox{f}}_Y. 
\end{eqnarray} 
From the counterpart of this equation that applies when only $X$ or only $Y$ is measured, we read off that the relative state of Alice and Bob with respect to Alice's external observer in state
$\ket{\mbox{o}}_X$ is $\ket{\mbox{ok}}_{A}\ket{1}_{B}$; with respect to Bob's external observer in state $\ket{\mbox{o}}_Y$ it is 
$\ket{h}_{A}\ket{\mbox{ok}}_{B}$ (in both cases only one external measurement is considered to take place, either $X$ or $Y$). These results are in accordance with what was stated in the above ``inconsistency argument''.

However, from Eq.\ 5 the state of Alice and Bob relative to the \emph{combined} external observers state $\ket{\mbox{o}}_X \ket{\mbox{o}}_Y$ is:
\begin{equation}
\ket{\mbox{ok}}_{A}\ket{\mbox{ok}}_{B} = \frac{1}{2} (\ket{h}_A - \ket{t}_A) (\ket{0}_B - \ket{1}_B).
\end{equation}
This is an entangled state in which neither the coin toss nor the qubit measurement has a definite result---it is \emph{not} the state $\ket{h}\ket{1}$ that was argued to be present in Bub's version of the F-R argument. This illustrates the fact that in the case of an entangled state between two systems, the perspectives of an external observer who measures one system and the observer who measures the other can generally not be glued together to give us the perspective of the system that consists of both observers. In fact, as we see from Eq.\ 5, in the final quantum state not only Alice and Bob, but also the external observers have become entangled with each other---this should already make us  suspicious of combining partial viewpoints into a whole, as it is well known that entanglement may entail non-classical holistic features.

So perspectival views, which make the assignment of properties dependent on the relative quantum state, are able to escape the above inconsistency argument by denying that the $X$-perspective and the $Y$-perspective can be simply juxtaposed to form the $XY$-perspective. 

A further point to note is that the quantities measured by the external observers do not commute with $A$ and $B$, respectively. So the measurement by the observer near to Alice introduces a ``context'' that is different from that of Alice's $A$-measurement, and similarly for the $Y$ and $B$ measurements. This makes it understandable that the combined final $XY$ perspective need not agree with the  measurement outcomes initially found by Alice and Bob \cite{fortin}. Indeed, the correct ``Alice and Bob state'' from the $XY$-perspective, given by Eq.\ 6, does not show one definite combination of results of Alice's $A$ and Bob's $B$ measurement but contains all of them as possibilities. 

The Frauchiger and Renner argument, in Bub's formulation, therefore does not threaten perspectival one-world interpretations with unitary dynamics. But we should wonder whether \emph{non-perspectival} unitary schemes, like the Bohm theory, will also be able to escape inconsistency; and if so, exactly how they do so. When we again follow the measurement steps in the thought experiment, we are allowed to conclude that the $X$ result ``ok'' can only be found if Bob had measured ``$1$''. In the context of  the Bohm interpretation this means that after the first external measurement the particle configuration of $X$ is at a point of configuration space that is compatible with the state $\ket{\mbox{o}}_X$ only if Bob's configuration is in a part of configuration space  compatible with $\ket{1}_B$. The same conclusion can be drawn with respect to $Y$ and $A$ for a measurement series in which $Y$ is measured before $X$: if the $Y$ measurement is performed first and  the result ``ok'' is registered, $A$ must have seen ``heads''. Now, $X$ and $Y$ are at space-like separation from each other, and this seems to imply that it cannot make a difference to the initial state of Alice and Bob what the order in time is of the $X$ and $Y$ measurements. If the $X$ measurement is the earlier one, Bob must have been in state $\ket{1}_B$; if $Y$ is measured first, Alice was in state $\ket{h}_A$. Therefore, if the time order is immaterial, Alice and Bob together were initially with certainty in the state $\{h,1\}$, if the $X,Y$ measurements ended with the result $\{\mbox{ok},\mbox{ok}\}$. But this is in contradiction with what the unitary formalism predicts: inspection of the initial state (Eq.\ (1)) shows that the outcome combination $\{h,1\}$ is impossible.  So we have an inconsistency, and the Bohm interpretation, and other non-perspectival unitary interpretations, seem to be in trouble.        

However, in the Bohm theory the existence of a preferred reference frame that defines a universal time is assumed to exist (see for more on the justification of this assumption section \ref{relativity}). This makes it possible to discuss the stages of the experiment in their objectively correct temporal order. Assume that after Alice's and Bob's ``in-the-room'' measurements (the first by Alice, the second by Bob), the external observer near to Alice measures first, after which the observer close to Bob performs the second measurement. The first external measurement will disturb the configuration of particles making up Alice and her environment, so Alice's initial result will be changed. If the result of the $X$ measurement is ``ok'', Bob's internal result must be $1$ (Bob is far away, but the possibility of this inference is not strange, because the total state is entangled, which entails correlations between Alice and Bob). This $1$ will  remain unchanged until the second external measurement, of $Y$. This second measurement will change Bob's state. Now, in this story it is \emph{not} true that the outcome ``ok'' of $Y$ is only compatible with Alice's outcome ``heads'', as used in the inconsistency argument: the previous measurement of $X$ has blocked this conclusion. In fact, the initial outcome combination $\{t,1\}$ will lead to the later results $\{\mbox{ok},\mbox{ok}\}$ with probability $1/4$, and since $\{t,1\}$ itself has probability $1/3$ in the initial state this leads to the correct probability of $1/12$ of $\{\mbox{ok},\mbox{ok}\}$. So there is a fully consistent story here, in which Alice's result is $t$.

So also Bohm and possibly other non-perspectival schemes are able to escape the inconsistency argument. In the perspectival schemes the key was that two perspectives cannot always be simply combined into one global perspective; because of this, we were allowed to speak about the $X$ perspective and the $Y$ perspective without specifying the temporal order of the $X$ and $Y$ measurements. The threat of inconsistency was avoided by blocking the composition of the two perspectives into one whole. In the non-perspectival scheme the existence of a preferred frame of reference comes to the rescue and protects us against inconsistency: we can follow the interactions and the changes produced by them step by step in their unique real time order, so that no ambiguity arises about which measurement comes first and about what the actual configuration is at each instant. The issue of combining descriptions from different perspectives accordingly does not arise, and this is enough to tell one consistent story.

The difference between the perspectival and non-perspectival unitary accounts, and the apparent connection between  perspectivalism and Lorentz invariance, suggests that there is a link between perspectivalism and relativity. Perspectivalism seems able to avoid inconsistencies without introducing a privileged frame of reference. On the other hand, the introduction of such a privileged frame in Bohm-like interpretations now appears as a ploy to eliminate the threat of inconsistencies without adopting perspectivalism.

\section{Relativity}\label{relativity}

The diagnosis of the previous section is confirmed when we directly study the consequences of relativity for interpretations of quantum mechanics. In particular, when we attempt to combine special relativity with unitary interpretational schemes, new hints of perspectivalism emerge. As mentioned, the Bohm interpretation has difficulties in accommodating Lorentz invariance. Bohmians have therefore generally accepted the existence of a preferred inertial frame in which the equations assume their standard form---a  frame resembling the ether frame of prerelativistic electrodynamics. Accepting such a privileged frame in the context of what we know about special relativity and Minkowski spacetime is of course not something to be done lightly; it must be a response to a problem of principle. Indeed, it can be mathematically proved that \emph{no} unitary interpretation scheme that attributes always definite positions to particles (as in the Bohm theory) can satisfy the requirement that the same probability rules apply equally to all hyperplanes in Minkowski spacetime \cite{berndl}. 

The idea of this  theorem (and of similar proofs) is that intersecting hyperplanes should carry properties and probabilities in a coherent way, which means that they should give agreeing verdicts about the physical conditions at the spacetime points where they (the hyperplanes) intersect. The proofs demonstrate that this meshing of hyperplanes is impossible to achieve with properties that are hyperplane independent. The no-go results can be generalized to encompass non-perspectival unitary interpretations that attribute other definite properties than position, and to unitary interpretations that work with sets of properties that change in time \cite{dickson}.  A general proof along these lines was given by Myrvold \cite{myrvold}. Myrvold shows, for the case of two systems that are (approximately) localized during some time interval, that it is impossible to have a joint probability distribution of definite properties along four intersecting hyperplanes such that this joint distribution returns the Born probabilities on each hyperplane. An essential assumption in the proof \cite[p.\ 1777]{myrvold} is that the properties of the considered systems are what Myrvold calls \emph{local}: the value of quantity $A$ of system $S$ at spacetime point $p$ (a point lying on more than one hyperplanes) must be well defined regardless of the hyperplane to which $p$ is taken to belong and regardless of which other systems are present in the universe. It turns out that such local properties cannot obey the Born probability rule on each and every hyperplane. The assumption that the Born rule only holds in a preferred frame of reference is one way of responding to this no-go result. 

The argument has been given a new twist by Leegwater \cite{leegwater}, who argues that ``unitary single-outcome quantum mechanics'' cannot be ``relativistic'', where a theory is called relativistic if all inertial systems have the same status with respect to the formulation of the dynamic equations of the theory (i.e., what usually is called Lorentz or relativistic invariance). Like Frauchiger and Renner, Leegwater considers a variation on the Wigner's friend thought experiment: there are three laboratory rooms, at spacelike distances from each other, each with a friend inside and a Wigner-like observer outside. In each of the lab rooms there is also a spin-$1/2$ particle, and the experiment starts in a state in which the three particles (one in each room) have been prepared in a so-called GHZ-state \cite{greenberger}. The description of the thought experiment in an initially chosen inertial rest frame is assumed to be as follows: at a certain instant the three friends inside their respective rooms simultaneously measure the spins of their  particles, in a certain direction; thereafter, at a second instant, each of the three outside observers performs a measurement on his room. This measurement is of a ``whole-room'' observable, like in the Frauchiger-Renner thought experiment discussed in the previous section. As a result of the internal measurements by the three friends the whole system consisting of the  rooms and their contents has ended up in an entangled GHZ-state. Leegwater is able to show that this entails that the assumption that the standard rules of quantum mechanics apply to each of three differently chosen sets of simultaneity hyperplanes, gives rise to a GHZ-contradiction: the different possible measurement outcomes (all $+1$ or $-1$) cannot be consistently chosen such that each measurement has the same outcome irrespective of the simultaneity hyperplane on which it is considered to be situated (and so that all hyperplanes mesh). As in the original GHZ-argument \cite{greenberger}, the contradiction is algebraic and does not involve the violation of probabilistic (Bell) inequalities.

One way of responding to these results is the introduction of a preferred inertial system (a privileged perspective!), corresponding to a state of absolute rest, perhaps defined with respect to an ether. This response is certainly against the spirit of special relativity, in particular because the macroscopic predictions of quantum mechanics are such that they make the preferred frame undetectable. Although this violation of relativistic invariance does not constitute an inconsistency, it certainly is attractive to investigate whether there exist other routes to escape the no-go theorems. Now, as we have seen, a crucial assumption in these theorems is that properties of systems are monadic, independent of the presence of other systems and independent of the hyperplane on which they are considered. This suggests that a transition to relational or perspectival properties offers an alternative way out.

In fact, that unitary evolution in Minkowski spacetime leads naturally to a hyperplane dependent account of quantum states if one wants to describe measurements by effective global collapses has been noted in the literature before (see e.g.\ \cite{diekscov,fleming,myrvold2}). The new light that we propose to cast on these and similar results comes from not thinking in terms of collapses, and of a dependence on hyperplanes or foliations of Minkowski spacetime \emph{as such}, but instead of interpreting them as consequences of the perspectival character of physical properties: that the properties of a system are defined with respect to other systems. What we take the considerations in the previous and present sections to suggest is that it makes a difference whether we view the physical properties of a system from one or another system; or from one or another temporal stage in the evolution of a system. In the case of the (more or less) localized systems that figure in the relativistic no-go theorems that we briefly discussed, this automatically leads to property ascriptions that are different on the various hyperplanes that are considered. As a result, the meshing conditions on which the theorems hinge no longer apply.

\section{Concluding Remarks}\label{conclusion}    
    
If unitary evolution is accepted as basic in quantum mechanics, and is combined with the requirement that results of experiments are definite and unique, this naturally leads to a picture in which physical systems have properties that are relational or perspective dependent. As we have seen in section \ref{context}, perspectivalism makes it possible to escape arguments saying that interpretations of unitary quantum mechanics in terms of one single world are not possible. Moreover, perspectivalism removes obstacles to the possibility of formulating interpretational schemes that respect Lorentz invariance by making the introduction of a preferred inertial frame of reference superfluous (section \ref{relativity}).

The single world that results from perspectivalism is evidently much more complicated than the world we are used to in classical physics: there are more than one valid descriptions of what we usually think of as one physical situation. This reminds of the many-worlds interpretation. There are important differences, though, between a multiplicity of worlds and the multiplicity of descriptions in perspectivalism. According to the single world perspectivalism that we have sketched only one of the initially possible results of a measurement becomes actual from the perspective of the observer, whereas in the many-worlds interpretation all possibilities are equally realized. So the multiplication of realities that takes place in many-worlds is avoided in perspectivalism. It is of course true that perspectivalism sports a multiplicity of its own, namely of different points of view within a single world. But this multiplicity seems unavoidable in the many-worlds interpretation as well, in each individual branch. For example, in the relativistic meshing argument of Myrvold \cite{myrvold}, a situation is discussed in which no measurements occur: the argument is about two freely evolving localized systems as described  from a number of different inertial frames. Since no measurements are taking place during the considered process, the inconsistency argument goes through in exactly the same way  in every single branch of the many-worlds super-universe: there is no splitting during the time interval considered in the proof of the theorem. So even in the many-worlds interpretation the introduction of perspectival properties (in each single branch) seems unavoidable in order to avoid inconsistencies. Another case to be considered is the Wigner's friend experiment: when the measurement in the hermetically sealed room has been performed, an outside observer will still have to work with the superposition of the two branches. So the splitting of worlds assumed  by the many-worlds interpretation must remain confined to the interior of the room, as mentioned in section \ref{noncollapse}. In this situation it is natural to make the description of the measurement and its result perspective dependent: for the friend and his copy inside the room there is a definite outcome, but this is not so for the external observer. So perspectivalism as a consequence of holding fast to unitarity and Lorentz invariance seems more basic than the further choice of interpreting measurements in terms of many worlds; even the many-worlds interpretation must be committed to perspectivalism. But perspectivalism on its own is already sufficient to evade the anti single-world arguments of section \ref{context}, so for this purpose we do not need the further assumption of many worlds.

Finally, the introduction of perspectivalism opens the door to several new questions. In everyday circumstances we do not notice consequences of perspectivalism, so we need an account of how perspectival effects are washed out in the classical limit. It is to be expected that decoherence plays an important role here, as alluded to in the Introduction---however, this has to be further worked out (cf.\ \cite{benedieks}). Further, there is the question of how the different perspectives hang together; for example, in section \ref{context} it was shown that perspectives of distant observers cannot be simply combined in the case of entanglement, which may be seen as a non-local aspect of perspectivalism. By contrast, it has been suggested in the literature that perspectivalism makes it possible to give a purely local description of events in situations of the Einstein-Podolsky-Rosen type, and several tentative proposals have been made in order to substantiate this \cite{rovelli,rovelli2,laudisa,benedieks,dieksrel}. These and other questions constitute largely uncharted territory that needs further exploration.

\end{document}